\documentclass[reqno]{amsart}
\usepackage{graphicx}
\newcommand{\abs}[1]{\left\vert#1\right\vert}
\newcommand{\vacuum}{|0\rangle}
\newcommand{\alphaprime}{\alpha '}
\newcommand{\zbar}{\overline{z}}
\DeclareMathOperator{\re}{Re} \DeclareMathOperator{\im}{Im}
\DeclareMathOperator{\Tr}{Tr}
\numberwithin{equation}{section}
\begin{document}

\hfill {HU-EP-03/80}\\ \vspace{0.1cm} \hfill DAMTP-2003-130\\
\vspace{0.1cm}
\hfill {\tt hep-th/0311182}\\

\vspace{1cm}

\begin{center}
{\large \bf Tachyon Condensation for Intersecting Branes at Small and
Large Angles}
\end{center}
\vskip1cm

\begin{center}
{\bf Friedel Epple$^a$ and Dieter L\"ust$^b$}
\end{center}

\vskip0.5cm
\begin{center}
{\sl
${}^a$Centre for Mathematical Sciences

Wilberforce Road

 Cambridge CB3 0WA, UK
}

email: F.Epple@damtp.cam.ac.uk
\end{center}

\vskip0.3cm
\begin{center}
{\sl
${}^b$ Humboldt-Universit\"at zu Berlin, Institut f\"ur Physik

Newtonstra{\ss}e 15

 D-12489 Berlin, Germany}

email: luest@physik.hu-berlin.de
\end{center}


\vskip1cm
\begin{center}
ABSTRACT
\end{center}
\vskip0.3cm We review the worldsheet analysis for intersecting
branes with focus on small and large angles. For small angles, we
review the Yang-Mills fluctuation analysis in ref.
\cite{Hashimoto03} and find an additional family of massless
modes. They are the components of a Goldstone scalar corresponding
to the spontaneously broken $U(2)$-gauge symmetry. For branes at
large angles, we derive an effective tachyon field theory from
BSFT results. We show how the gauge symmetry of this system
implies a mass spectrum which is consistent with the worldsheet
analysis.

\tableofcontents

\section{Introduction}
Intersecting branes have been at the centre of many recent
developments in brane cosmology and string phenomenology. In
particular, they have been used to construct realistic brane world
models which reproduce the Standard Model particle spectrum at low
energies \cite{Blumenhagen:2000wh, Angelantonj:2000hi,
Aldazabal:2000dg, Aldazabal:2000cn, Blumenhagen:2000ea,
Ibanez:2001nd,Forste:2001gb, Blumenhagen:2001te,
Cvetic:2001tj,Cvetic:2001nr,Bailin:2001ie, Kokorelis:2002zz} (a
more complete list on intersecting brane world models can be e.g.
found in \cite{Blumenhagen:2003qd}). In particular, intersecting
brane worlds provide a natural explanation for family replication
and Yukawa coupling hierarchy
\cite{Ibanez:2001nd,Cvetic:2002wh,Cremades:2003qj}. Intriguingly,
the Standard Model Higgs effect might be realized by tachyon
condensation \cite{Aldazabal:2000dg,Blumenhagen:2001te} which is
widely believed to trigger brane recombination and correspondingly
a reduction of the gauge group's rank. In cosmology, intersecting
brane worlds have been used to model early universe inflation and
a ``graceful exit'' from the inflationary period
\cite{Alexander:2001ks,Burgess:2001fx,Dvali:2001fw,
Shiu:2001sy,Herdeiro:2001zb,
Burgess:2001vr,Garcia-Bellido:2001ky,Blumenhagen:2002ua,
Padmanabhan02, Gomez-Reino02}.

Although it is well known that intersecting branes are generically
unstable, the focus has mainly been on static properties of fixed
brane configurations rather than on dynamical aspects. Where brane
dynamics were considered - such as in brane recombination
processes - the arguments have been mostly based on considering
conserved D-brane charges rather the full tachyon potential. (The
nature of supersymmetric D-brane bound states for intersecting
branes after tachyon condensation was described in
\cite{Blumenhagen:2000eb}, a world volume perspective on the
recombination of intersecting branes was given in \cite{Erdmenger}
and some other relevant D-brane bound states were discussed in
\cite{Ohta}.) The reason is of course that a quantitative
description of brane dynamics would in principle have to account
for all of the infinitely many string modes. This is clearly very
difficult. However, in many cases the full string interactions can
be truncated to give an effective field theory description.

Various field theory actions of multi-brane dynamics have been
proposed with different regimes of validity and some of them have
proven to be simple but powerful tools (see also
figure 1). While an effective field
theory would generally be expected to reproduce relevant parts of
the string theory mass spectrum and scattering amplitudes, some
field theory models go far beyond this minimum requirement.
Starting from a BPS configuration, e.g. two parallel D-branes, and
then rotating them by a small angle $\theta$, one single tachyon
field shows up in the open string spectrum. Hence in this case an
effective field theory with a finite number of fields  certainly
provides for an appropriate description of the tachyon dynamics.
However, when the intersecting angle is growing, more and more
string modes will become tachyonic. In particular settings where
the intersection angle $\theta$ is close to $\pi$, which just
corresponds to the small angle intersection of a brane-antibrane
system, contain a large number of tachyonic modes, as we will see
from a world-sheet analysis. For the case $\theta=\pi$ the
infinitely many tachyons become tachyonic momentum states, and the
corresponding coincident brane-antibrane pair is a highly unstable
non-BPS state, and does not correspond to a perturbative string
ground state. Nevertheless, a number of essentially
non-perturbative phenomena have been realized on a field theory
level, notably brane descent relations \cite{Minahan00b,
Hashimoto02, Sen03}, decay of non-BPS branes \cite{Gibbons00,
Sen02b}, brane-antibrane annihilation \cite{Minahan00b, Sen02c}
and local brane recombination \cite{Hashimoto03}. It is one of the
aims of this paper to generalize these effective potentials to
cases where branes and antibranes intersect each other at a small
angle, i.e. the large angle case of intersecting brane.

A convenient way to study the dynamics of intersecting branes is
provided by switching to the equivalent T-dual picture, where the
geometrical intersection angle is transformed to an open string
gauge field strength background on the D-branes. The T-dual
picture is very useful, since non-Abelian Yang-Mills theory is
probably the best known approximation to multi-brane dynamics. For
small angles, it has been shown that its fluctuation mass spectrum
agrees with the string theory worldsheet analysis
\cite{Hashimoto97}. Also, brane recombination via tachyon
condensation has recently been realized within the framework of
Yang-Mills theory. However, one should keep in mind that because
of the ``slowly-varying field approximation'', the Non-Abelian
Yang-Mills description of brane dynamics is in principle only
valid for small intersection angles.

For the Abelian case, the Yang-Mills action is the first term in
an expansion of the Born-Infeld action. The Born-Infeld action in
turn is a valid truncation of the full string theory brane
dynamics for any constant magnetic flux (by T-duality, this
corresponds to constant brane slope). The situation for a
multi-brane background is much more difficult as there are grave
problems in finding an equivalent to the Born-Infeld action in the
Non-Abelian case. At the core of the problem are ordering
ambiguities which are caused by non-commutativity. In 1997,
Tseytlin proposed a symmetrized trace prescription
\cite{Tseytlin97} which subsequently was shown to reproduce the
string theory spectrum on intersecting branes up to $F^4$-accuracy
\cite{Hashimoto97}. Higher order corrections were derived
\cite{Koerber01, Koerber02} and recently tested by comparing their
fluctuation spectrum to the results from string theory worldsheet
analysis \cite{Sevrin03, Nagaoka03}. In addition, in the context
of the heterotic/type I string duality parts of the Non-Abelian
Born-Infeld action were computed by direct computation of string
scattering amplitudes \cite{Stieberger:2002fh,Stieberger:2002wk}.
However, although these recent versions of the Non-Abelian
Born-Infeld action do have a higher accuracy than the Non-Abelian
Yang-Mills approximation, their range of validity is still limited
on principle to small field strength magnitude (and small angles
in the T-dual perspective). Hence, the difficulties in finding an
effective action for general field strength magnitude mean that we
really do not have a valid field theory description of string
theory in the presence of branes intersecting at large angles.

\begin{figure}
  \centering
  \includegraphics[width=12cm]{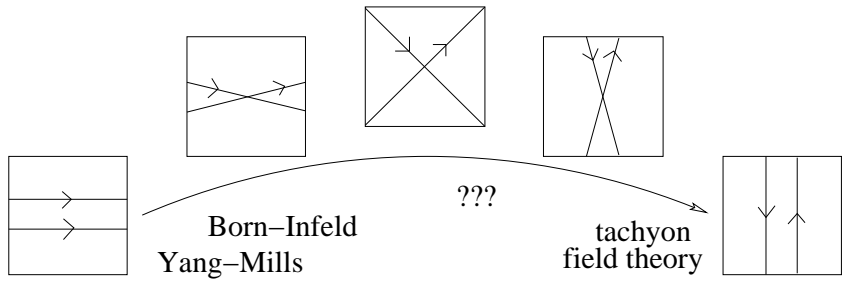}
  \caption{Intersecting D-branes: For a small intersection angle, the Non-Abelian Yang-Mills action is a valid
  field theory description. The Non-Abelian Born-Infeld action is more accurate but in principle also valid only
  for a small intersection angle. While a field theory description at large angle is not known,
  tachyon field theory models have been proposed for the brane-antibrane system.}\label{YMtoTEFT}
\end{figure}

On the other hand, there have been a number of exciting
discoveries in string field theory which eventually led to
effective field theory descriptions of the brane-antibrane system
\cite{Minahan00b, Sen02c, Sen03}. From the perspective of
intersecting branes, the brane-antibrane setting is obtained by
taking one intersection angle to its maximum value. It is
therefore natural to ask whether one can generalize the
brane-antibrane field theory actions to include settings where a
brane and an antibrane intersect at small angles. This would be
the equivalent to Yang-Mills theory in the brane-antibrane case.
Figure \ref{YMtoTEFT} shows a graphic depiction of the above
discussion.

As a preparation for later arguments, we review the worldsheet
analysis of branes at angles in section \ref{SecWS}. We will see
that for small angles there is just one tachyonic field, a fact
which naturally allows for a straightforward effective field
theory treatment of the tachyon condensation. In contrast, the
case of large angles is much more subtle since it contains a
growing number of tachyonic field, and stringy methods are
required for a proper description. In section \ref{SecFT} we
shortly review the Yang-Mills description of branes at small
angles with respect to mass spectrum matching and the role of
tachyon condensation. Then, we examine brane-antibrane effective
field theories and generalize these to configurations where brane
and antibrane intersect at small angles. This is a first step
towards filling the gap between branes at small angles and the
parallel brane-antibrane case. In particular, we establish an
effective action for the intersecting brane-antibrane pair by
gauging a well known tachyon action for the parallel
brane-antibrane pair. We argue that the specific gauge symmetry of
this background ensures that the field theory fluctuation spectrum
is consistent with the string theory worldsheet analysis.


\section{Worldsheet analysis of branes at angles}
\label{SecWS}

Branes at angles have been discussed from a string worldsheet
perspective in a number of papers \cite{Berkooz96, Arfaei96,
Jabbari97, Hashimoto97,Blumenhagen:2000fp}.
However, these discussions have focussed
on special geometries and/or small angles. To prepare the stage
for the following sections, we will review the worldsheet analysis
of branes at angles. Although we will follow roughly the
discussion in \cite{Berkooz96, Hashimoto97}, we use different
conventions and notation in order to keep the discussion fully
general and transparent at the same time.

The boundary conditions for a string connecting the two branes at
angles can be written in a compact way by introducing complex
coordinates $Z^{1}=X^{2}+iX^{3}$, $Z^2=X^4+iX^5$, $\dots$ and
similarly for the worldsheet fermions.\footnote{The coordinate
$X^1$ is usually reserved for an optional non-zero separation of
the branes.} This allows us to express the rotation taking the
first brane to the second one by $Z^{a}\rightarrow
\exp{-i\theta_{i}} Z^{a}$. Using this notation, the boundary
conditions become:

\begin{eqnarray}
\partial_{\sigma} \re(Z^{a})|_{\sigma=0}&=& 0 \\
\im(Z^{a})|_{\sigma=0}&=&0 \\
\partial_{\sigma} \re[\exp(i\theta_{a})Z^{a}]|_{\sigma=\pi}&=&0 \\
\im[\exp(i\theta_{a})Z^{a}]|_{\sigma=\pi}&=&0
\end{eqnarray}

for the complex bosonic field. For the following considerations,
it is convenient to measure angles in units of $\pi$. If we
introduce the quantities $\alpha_a\equiv\theta_a/\pi$, the
classical solutions to the string equations of motion with the
above boundary conditions have the following mode expansion

\begin{align}
Z^a(z,\zbar)&=i\left(\frac{\alphaprime}{2}\right)\sum_{n\in\mathbb{Z}}
\left\{\frac{x^{a}_{n-\alpha_a}}{(n-\alpha_a)z^{n-\alpha_a}}+\frac{x^{a\dagger}_{n+\alpha_a}}{(n+\alpha_a)\zbar^{n+\alpha_a}}\right\}\\
Z^{a\dagger}(z,\zbar)&=i\left(\frac{\alphaprime}{2}\right)\sum_{n\in\mathbb{Z}}
\left\{\frac{x^{a\dagger}_{n+\alpha_a}}{(n+\alpha_a)z^{n+\alpha_a}}+\frac{x^{a}_{n-\alpha_a}}{(n-\alpha_a)\zbar^{n-\alpha_a}}\right\},
\end{align}

where $z=e^{i\sigma+\tau}$. Due to the complex nature of $Z$ we do
not have $x^{a\dagger}_{m-\alpha_a}=x^a_{-m+\alpha}$. From
canonical quantization, we get the following commutator relations

\begin{equation}
[x^{a\dagger}_{-m+\alpha_a},x^{b}_{n-\alpha_b}]=(-m+\alpha_a)\delta_{ab}\delta_{mn}.
\end{equation}

  From worldsheet supersymmetry, it follows that the complexified
worldsheet fer\-mions have the same moding as the worldsheet bosons
(Ramond fermions) or an additional shift by $1/2$ (Neveu-Schwarz
fermions). In the following discussion, we will focus on the NS
sector only because this is where tachyonic modes appear. The mode
expansion becomes

\begin{equation}
\Psi^a(z)=\sum_{r\in\mathbb{Z}+\frac{1}{2}}\frac{{\psi^a_{r-\alpha_a}}}{z^{r-\frac{1}{2}-\alpha_a}},\qquad
\Psi^{a\dagger}(z)=\sum_{s\in\mathbb{Z}+\frac{1}{2}}\frac{\psi^{a\dagger}_{s+\alpha_a}}{z^{s+\frac{1}{2}+\alpha_a}},
\end{equation}

where the standard doubling trick has been used to extend the
parameter range of the fermions to $\sigma\in[0,2\pi)$. The
canonical anti-commutation relations are given by

\begin{equation}
\{\psi^{a\dagger}_{-r+\alpha_a},\psi^{b}_{s-\alpha_b}\}=\delta_{ab}\delta_{rs}.
\end{equation}

The Virasoro generator $L_0$ is given by

\begin{equation}
L_0=\sum_a\left\{\sum_{m\in\mathbb{Z}}:x^{a\dagger}_{-m+\alpha_a}x^a_{m-\alpha_a}:
+\sum_{r\in\mathbb{Z}+\frac{1}{2}}(r-\alpha):\psi^{a\dagger}_{-r+\alpha}\psi^a_{r-\alpha}:\right\}+\varepsilon_0,\label{normalordering}
\end{equation}

with a zero point energy $\varepsilon_0$ which can be computed
from $\zeta$-function regularization of the infinite sums which
arise in the process of normal ordering $L_0$. The exact meaning
of normal ordering depends on the definition of the vacuum state.
Note that the classification of mode operators in terms of
creation and annihilation operators is essentially arbitrary.
However, shifting an operator from ``creators'' to
``annihilators'' has an effect on the normal ordering constant
such that the mass spectrum of the theory is unaffected. We will
return to this issue when we discuss negative angles,
$\alpha_a<0$. For positive angles, $\alpha_a>0$, it is appropriate
to define the vacuum state by the properties

\begin{align}
x^{a}_{m-\alpha_a}\vacuum &=0,\quad m\geq 1 \\
x^{a\dagger}_{-m+\alpha_a}\vacuum &=0,\quad m\leq 0 \\
\psi^a_{r-\alpha_a}\vacuum &=0, \quad r\geq 1/2 \\
\psi^{a\dagger}_{-r+\alpha_a}\vacuum &=0,\quad r\leq -1/2.
\end{align}

With respect to this definition, the normal ordering constant
$\varepsilon_0$ can be computed by a zeta function regularization
procedure. The contribution due to the $a$'th complex worldsheet
fields is given by (here, we are suppressing the superscript $a$
in the worldsheet fields):

\begin{align}
\varepsilon_0^{(a)}&=\sum_{m=-\infty}^0[x^{\dagger}_{-m+\alpha},x_{m-\alpha}]
+\sum_{r=-\infty}^{-1/2}(r-\alpha)\{\psi^{\dagger}_{-r+\alpha},\psi_{r-\alpha}\}\nonumber\\
&=\sum_{m=-\infty}^0(-m+\alpha)+\sum_{m=-\infty}^{-1/2}(r-\alpha)\nonumber\\
&=\sum_{m=0}^{\infty}(m+\alpha)-\sum_{m=0}^{+\infty}(m+\frac{1}{2}+\alpha)\nonumber\\
&=\zeta(-1,\alpha)-\zeta(-1,1/2+\alpha)
\end{align}

where $\zeta$ denotes the generalized or Hurwitz zeta function
which is defined by

\begin{equation}
\zeta(s,a)=\sum_{k=0}^{\infty}\frac{1}{(k+a)^s}.
\end{equation}

The Hurwitz zeta function reduces to the ordinary (Riemann) zeta
function for $a=0$. It has a well-defined analytical continuation
to $s=-1$. To complete the computation of the zero point energy,
we use a well-known property of the Hurwitz zeta function, which
is given by:

\begin{equation}
\zeta(-1,\alpha)=-\frac{1}{12}+\frac{1}{2}\alpha-\frac{1}{2}\alpha^2
\end{equation}

>From this, it follows that the contribution from the $a$'th
complex fields becomes

\begin{equation}
\varepsilon_0^{(a)}=-\frac{1}{8}+\frac{1}{2}\alpha_a.
\end{equation}

In the computation of the total zero point energy, one has to take
into account the Fadeev-Popov ghosts. As usual, their contribution
cancels the contribution of two components of the worldsheet
fields. There remain eight real dimensions, which add up to a
value

\begin{equation}
\varepsilon_0=-\frac{1}{2}+\frac{1}{2}\sum_{a=1}^{4}\alpha_a.
\end{equation}

Before we move on to determining the low-energy mass spectrum of
branes at angles, we consider negative angles, $\alpha_a<0$. In
this case, it turns out that it is most convenient to work with a
slightly different definition of the vacuum state, namely:

\begin{align}
x^{a}_{m-\alpha_a}\vacuum &=0,\quad m\geq 0 \\
x^{a\dagger}_{-m+\alpha_a}\vacuum &=0,\quad m\leq -1 \\
\psi^a_{r-\alpha_a}\vacuum &=0, \quad r\geq 1/2 \\
\psi^{a\dagger}_{-r+\alpha_a}\vacuum &=0,\quad r\leq -1/2.
\end{align}

This definition differs from the one we used for positive angles
in that $x^a_{-\alpha_a}$ is an annihilation operator while
$x^{a\dagger}_{\alpha_a}$ is a creation operator. As a consequence
of this shift, the zero point energy picks up an additional term
of the form $-\alpha_a$. Then, the total zero point energy can be
written in unified way

\begin{equation}
\varepsilon_0=-\frac{1}{2}+\frac{1}{2}\sum_{a=1}^4 \abs{\alpha_a},
\end{equation}

which is valid for both positive and negative angles. The vacuum
state itself is removed by a generalization of the GSO-projection
\cite{Arfaei96}. Any physical state has to contain at least one
fermionic creation operator.


\subsection{D-branes at one angle}

As a simple example, we consider branes at one angle. These could
be intersecting D-strings or more generally a pair of Dp-branes
which intersect on a (p-1)-dimensional hyperplane. The latter
configurations are distinguished only by the number of additional
(and mutually perpendicular) Dirichlet directions. These
directions do not affect the mass spectrum. Thus, all such
configurations can be treated along the lines of the preceding
discussion by setting all angles but one to zero. The zero point
energy is then given by

\begin{equation}
\label{OneAngleZPE}
\varepsilon_0=-\frac{1}{2}+\frac{1}{2}\abs{\alpha}
\end{equation}

where $\alpha\equiv\alpha_1$ is defined by the single
non-vanishing angle. For small \emph{positive} $\alpha$, the
lowest mass state is

\begin{equation}
\psi^{\dagger}_{-1/2+\alpha}\vacuum
\end{equation}

whose mass is given by [cf. the expression \eqref{normalordering}
for $L_0$]:

\begin{equation}
\alphaprime
m^2=\varepsilon_0+\left(\frac{1}{2}-\alpha\right)=-\frac{\alpha}{2}.
\end{equation}

There is a tower of evenly spaced states,
$(x_{-\alpha})^n\psi^{\dagger}_{-1/2+\alpha}\vacuum$, with masses
$\alphaprime m^2=(-1/2+n)\alpha$, which builds upon the lowest
mass state. On the other hand, for small \emph{negative} $\alpha$,
there is a tower of states,
$(x^{\dagger}_{\alpha})^n\psi_{-1/2-\alpha}$, with masses
$\alphaprime m^2=(-1/2+n)(-\alpha)$. Both cases can be treated in
a unified way by writing

\begin{equation}
m^2=\left(-\frac{1}{2}+n\right)\abs{\frac{\theta}{\pi\alphaprime}},
\qquad n=0,1,2,\dots \label{lowlyingstringspectrum}
\end{equation}

for the lowest mass states, where $\theta$ is the intersection
angle.
This part of the mass spectrum is reproduced
to order $\theta$ by the spectrum of fluctuations around an
intersecting brane background in Non-Abelian Yang-Mills theory
with scalar fields \cite{Hashimoto03}. The T-dual configuration
which has first been discussed in \cite{Hashimoto97} is pure
Non-Abelian Yang-Mills theory around a background with non-zero
constant flux. The spectrum \eqref{lowlyingstringspectrum} has
also been used to check $F^4$- and higher order terms in the
Non-Abelian Born-Infeld action \cite{Hashimoto03, Nagaoka03}.

\begin{figure}[htb]
  \centering
  \includegraphics[width=12cm]{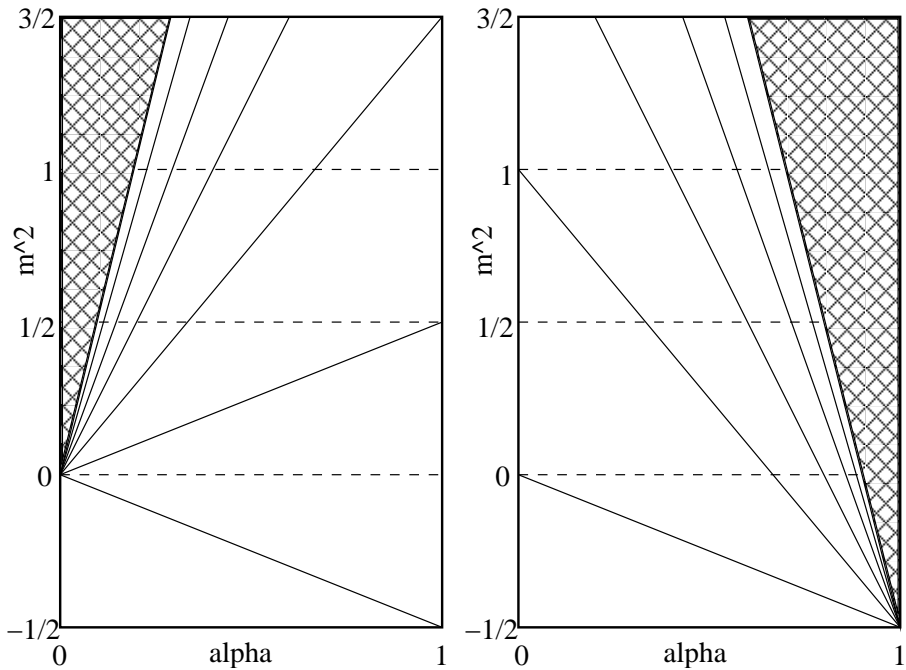}
  \caption{These plots show part of the NS spectrum on intersecting branes.
  The states on the left constitute the tower of lowest mass
states for configurations
  which are close to parallel branes.
These become momentum states at vanishing intersection angle ($\alpha=0$).
  The states on the
right become the (tachyonic) lowest mass states for configurations
  which are close to the brane-antibrane pair.
They become momentum states for $\alpha=1$.}
  \label{twotowers}
\end{figure}

However, it is important to note that this spectrum does no longer
represent the lowest mass excited states when one takes $\theta$
to be large.\footnote{Here, we discuss the case where $\theta>0$.
This can easily be modified to include negative angles} To see
what exactly happens to the low-lying mass spectrum when the
intersection angle flows from $\theta=0$ (parallel brane-brane
system) to $\theta=\pi$ (parallel brane-antibrane system), note
that the states created by $x^{\dagger}_{-\alpha}$ become
increasingly heavy while states created by $x_{-1+\alpha}$ become
increasingly light. At $\alpha=1/2$, which corresponds to
perpendicular branes, the two families of excitations exchange
roles. Finally, when one is approaching the brane-antibrane
system, the low-lying excitations are
$\psi^{\dagger}_{-1/2+\alpha} (x_{-1+\alpha})^{n}\vacuum$ and
their mass spectrum is given by

\begin{equation}
\alphaprime
m^2=\varepsilon_0+\left(\frac{1}{2}-\alpha\right)+n(1-\alpha),
\quad n=0,1,2,\dots
\end{equation}

Using the relation \eqref{OneAngleZPE} this becomes:

\begin{equation}
m^2=-\frac{1}{2\alphaprime}+\frac{\phi}{2\pi\alphaprime}(2n+1),
\quad n=0,1,2,\dots \label{DDbarspectrum}
\end{equation}

where $\phi\equiv\pi-\theta$ is the intersection angle
\emph{between the brane and antibrane}. In the limit of
$\phi\rightarrow 0$, the mass tower collapses and the
corresponding oscillators become momentum states. This is just
what happens for the brane-brane case, when $\theta\rightarrow 0$.
Figure \ref{twotowers} is a graphic representation of the
low-lying spectrum on intersecting branes. Note that due to our
definition of the vacuum state, $x_{-1/2+\alpha}$ is a creation
operator even for $\alpha>1/2$ in contrast to the usual
conventions. One could redefine the vacuum state for systems which
are close to the brane-antibrane system, thereby shifting
$x_{-1/2+\alpha}$ to the annihilation operators but this would
again imply a modification of the normal ordering constant which
ensures that the mass spectrum stays the same.


\subsection{D-branes at two and three angles}

As an illustration of the many-angles case, consider a pair of
intersecting D2-branes (alternatively: Dp-branes intersecting on a
(p-2)-dimensional hyperplane). For small positive intersection
angles, the lowest-lying states are given by

\begin{align}
\psi^{1\dagger}_{-1/2+\alpha_{1}}(x^{1}_{-\alpha_1})^{n_1}(x^{2}_{-\alpha_2})^{n_2}\vacuum
&& \quad \alphaprime m^2=(n_1-1/2)\alpha_1+(n_2+1/2)\alpha_2\nonumber\\
\psi^{2\dagger}_{-1/2+\alpha_{2}}(x^{1}_{-\alpha_1})^{n_1}(x^{2}_{-\alpha_2})^{n_2}\vacuum
&& \quad \alphaprime m^2=(n_1+1/2)\alpha_1+(n_2-1/2)\alpha_2\nonumber\\
\psi^1_{-1/2-\alpha_{1}}(x^{1}_{-\alpha_1})^{n_1}(x^{2}_{-\alpha_2})^{n_2}\vacuum
&& \quad \alphaprime m^2=(n_1+3/2)\alpha_1+(n_2+1/2)\alpha_2\label{D2D2spectrum}\\
\psi^2_{-1/2-\alpha_{2}}(x^{1}_{-\alpha_1})^{n_1}(x^{2}_{-\alpha_2})^{n_2}\vacuum
&& \quad \alphaprime m^2=(n_1+1/2)\alpha_1+(n_2+3/2)\alpha_2\nonumber\\
\psi^{\mu}_{-1/2}(x^{1}_{-\alpha_1})^{n_1}(x^{2}_{-\alpha_2})^{n_2}\vacuum
&& \quad \alphaprime
m^2=(n_1+1/2)\alpha_1+(n_2+1/2)\alpha_2\nonumber
\end{align}

where $\psi^{\mu}$ is a fermionic creation operator in a dimension
where both branes are point-like. The masses are computed by
adding the appropriate contributions from the creation operators
to the zero-point energy $\varepsilon_0$. Generically, the lowest
mass state is tachyonic with two evenly spaced towers of
low-energy excited states on top of it. The towers' spacing is
$\alpha_1$ and $\alpha_2$ respectively. There is one special
configuration where the lowest mass state becomes massless, namely
when $\theta_1=\theta_2$. This is a well-known supersymmetric
brane configuration. Contrastingly, if the angles have opposite
sign (say $\alpha_1>0$, $\alpha_2<0$), the lowest-lying states and
their masses are given by

\begin{align}
\psi^{1\dagger}_{-1/2+\alpha_{1}}(x^{1}_{-\alpha_1})^{n_1}(x^{2\dagger}_{\alpha_2})^{n_2}\vacuum
&& \quad \alphaprime m^2=(n_1-1/2)\alpha_1-(n_2+1/2)\alpha_2\nonumber\\
\psi^{2}_{-1/2-\alpha_{2}}(x^{1}_{-\alpha_1})^{n_1}(x^{2\dagger}_{\alpha_2})^{n_2}\vacuum
&& \quad \alphaprime m^2=(n_1+1/2)\alpha_1-(n_2-1/2)\alpha_2\nonumber\\
\psi^1_{-1/2-\alpha_{1}}(x^{1}_{-\alpha_1})^{n_1}(x^{2\dagger}_{-\alpha_2})^{n_2}\vacuum
&& \quad \alphaprime m^2=(n_1+3/2)\alpha_1-(n_2+1/2)\alpha_2\\
\psi^{2\dagger}_{-1/2+\alpha_{2}}(x^{1}_{-\alpha_1})^{n_1}(x^{2\dagger}_{-\alpha_2})^{n_2}\vacuum
&& \quad \alphaprime m^2=(n_1+1/2)\alpha_1-(n_2+3/2)\alpha_2\nonumber\\
\psi^{\mu}_{-1/2}(x^{1}_{-\alpha_1})^{n_1}(x^{2\dagger}_{-\alpha_2})^{n_2}\vacuum
&& \quad \alphaprime
m^2=(n_1+1/2)\alpha_1-(n_2+1/2)\alpha_2\nonumber
\end{align}

and the spectrum becomes tachyon-free for $\alpha_1=-\alpha_2$.
For branes at two small angles $\theta_1$ and $\theta_2$, the
lowest part of the mass spectrum can therefore be summarized by
the formula (not counting degeneracies)

\begin{equation}
\label{twoangles}
m^2=\left(-\frac{1}{2}+n_1\right)\abs{\frac{\theta_1}{\pi\alphaprime}}
+\left(-\frac{1}{2}+n_2\right)\abs{\frac{\theta_2}{\pi\alphaprime}},
\quad n_1,n_2=0,1,2,\dots
\end{equation}

Contrastingly, if for positive angles we let $\theta_2$ approach
$\pi$ while keeping $\theta_1$ small at the same time we get to a
configuration where a brane and an antibrane intersect at small
angles $\theta\equiv\theta_1$ and $\phi\equiv\pi-\theta_2$. The
lowest mass states are now given by

\begin{equation}
\psi^{2\dagger}_{-1/2+\alpha_2}(x_{-\alpha_1}^1)^{n_1}(x^2_{-1+\alpha_2})^{n_2}\vacuum
\end{equation}

and the corresponding mass spectrum is given by

\begin{equation}
m^2=-\frac{1}{2\alphaprime}+\frac{\theta}{\pi\alphaprime}(2n_1+1)
+\frac{\phi}{\pi\alphaprime}(2n_2+1), \quad n_1,n_2=0,1,2,\dots
\end{equation}

If we let both intersection angles approach their maximum value,
we once more end up with a brane-brane configuration because this
operation reverses the relative brane orientation twice.
Consistently, the mass spectrum is the same as in
\eqref{twoangles} upon substituting $\theta_1\rightarrow \pi
-\theta_1, \theta_2\rightarrow \pi-\theta_2$.

The discussion could now be easily continued also to the case of
D-branes at three angles. Here the tachyon-free, supersymmetric
configuration is obtained, if $\theta_1+\theta_2+\theta_3=\pi$.
There are four scalar fields living at each intersection that
could become tachyonic. The stability conditions are somewhat more
complicated compared to the case of two D-branes. In particular it
turns out there exist an extended region of stability, namely the
angle parameter space can be represented as a tetrahedron
\cite{Ibanez:2001nd,Rabadan:2001mt}. On the walls of the
tetrahedron one supersymmetry is preserved. For small deviations
outside the tetrahedron one of the four scalar fields becomes
tachyonic. So outside the tetrahedron the tachyon condensation
will takes place, and there will be a decay to another system.
Going to larger angles outside the tetrahedron one again enters
the region of getting more and more tachyonic fields. However
inside the tetrahedron there are no tachyons at all, and the
system is stable despite of not being supersymmetric.


\section{Branes at small angles}
\label{SecFT}

Before we consider field theory descriptions of branes at large
angles, we would like to review some results from the familiar
Yang-Mills action for branes at small angles such as its
fluctuation mass spectrum and the role of tachyon condensation.
The Yang-Mills description of low energy brane dynamics provides
an explicit realization of brane recombination processes and local
brane-antibrane annihilation. This has been shown recently in a
paper by Hashimoto and Nagaoka \cite{Hashimoto03}. They considered
a Yang-Mills background corresponding to intersecting D-Strings in
a non-compact geometry. An analysis of the Yang-Mills fluctuation
fields shows that the condensation of the tachyonic mode can be
related to brane-recombination by a local gauge transformation of
the brane-coordinates. The following paragraph will largely follow
their discussion of the fluctuation analysis, filling in some
further details. In particular, we compute the tachyon potential
in the Yang-Mills framework and discuss its relevance for brane
recombination processes. We also find an infinite family of moduli
in the fluctuation spectrum of the intersecting brane solution and
show that they generate gauge transformations of the background
fields.

The Yang-Mills action for the problem is the dimensionally reduced
pure Yang-Mills action

\begin{equation}
S=-\frac{1}{2}\int d^2x\Tr\left(F_{\mu\nu}F^{\mu\nu}+2D_{\mu}\Phi
D^{\mu}\Phi\right)\label{YMaction}
\end{equation}

where we choose to set the coupling constant to one and omit
fermions. The scalar field $\Phi$ is given by the scaled brane
coordinates in the $y$-direction: $\Phi=2\pi\alphaprime Y$. The
background solution corresponding to two D-Strings intersecting at
an angle $\theta$ is given by

\begin{gather}
\label{YMbackground}
\Phi=qx(1/2)\tau_3\\
A_{\mu}=0
\end{gather}

where q is related to the intersection angle by

\begin{equation}
\label{defofq} q=\frac{1}{\pi\alphaprime}\tan(\theta/2).
\end{equation}

Expanding the action (\ref{YMaction}) around the background
(\ref{YMbackground}) leads to the quadratic Lagrangian for the
fluctuation fields. The diagonal fluctuations (those which are
proportional to $\mathbf{1}$ or $\tau_3$) commute with the
background fields. Therefore they decouple from all other
fluctuations and satisfy free field equations. We can neglect them
in the following discussion. In order to follow the discussion in
\cite{Hashimoto03}, one has to impose one condition on the
fluctuation fields, namely $\delta A_0=0$. The remaining four
fluctuation fields are then grouped in pairs which decouple from
each other at the quadratic level. However, note that one cannot
simply use gauge invariance to go to Coulomb gauge because this
would at the same time modify the gauge field background. In fact,
choosing a specific background generally fixes the gauge symmetry
completely. K. Hashimoto, in private communication, justifies
setting $\delta A_0=0$ by saying that the gauge transformation
which is needed to do this is small when compared to the
background fields and that the small modification of the gauge
background could be interpreted as fluctuations in other sectors
than $A_0$ after the gauge transformation. We rather prefer to see
the spectrum one obtains by setting $\delta A_0=0$ as a
significant part of the full spectrum, which can be investigated
with relative ease. Having said this, we will from now on assume
$\delta A_0=0$ and continue in the analysis by defining:

\begin{gather}
\label{defofabphipsi}
\delta A_1^2\equiv a, \qquad\delta\Phi^1\equiv\varphi,\\
\delta A_1^1\equiv b, \qquad\delta\Phi^2\equiv\psi.
\end{gather}

Then, the quadratic Lagrangian for the fluctuation fields splits
in two parts: $\mathcal{L}=\mathcal{L}_A+\mathcal{L}_{B}$ with

\begin{align}
\label{fluctuationlagrangians}
\mathcal{L}_A&=\frac{1}{2}(\partial_t a)^2+\frac{1}{2}(\partial_t
\varphi)^2-\frac{1}{2}(\partial_x\varphi)^2-\frac{1}{2}q^2x^2a^2-qa\varphi+qxa\partial_x\varphi,\\
\mathcal{L}_B&=\frac{1}{2}(\partial_t b)^2+\frac{1}{2}(\partial_t
\psi)^2-\frac{1}{2}(\partial_x\psi)^2-\frac{1}{2}q^2x^2b^2+qb\psi-qxb\partial_x\psi.
\end{align}

We can decompose the fluctuation fields into their mass
eigenstates by setting

\begin{equation}
\begin{pmatrix}
  a(t,x) \\
  \varphi(t,x)
\end{pmatrix} = \sum_{n}
\begin{pmatrix}
  a_n(x) \\
  \varphi_n(x)
\end{pmatrix} \exp(-im_nt)
\end{equation}

and similarly for $(b,\psi)$. The mass spectrum is a priori
continuous but it will turn out that requiring normalizablility
for the wave functions (which is equivalent to demanding that the
action be finite) leads to an integer index $n$. The equations of
motion for the fluctuation fields $(a,\varphi)$ are

\begin{equation}
\begin{pmatrix}
  -q^2x^2+m_n^2 & +qx\partial_x-q \\
  -qx\partial_x-2q & \partial_x^2+m_n^2
\end{pmatrix}
\begin{pmatrix}
  a_n(x)\\
  \varphi_n(x)
\end{pmatrix}=0.
\label{matrixequation1}
\end{equation}

The equations of motion for $(b,\psi)$ can be derived from these
by substituting $(a,\varphi)\leftrightarrow (b,\psi)$ because this
substitution interchanges $\mathcal{L}_A$ and $\mathcal{L}_B$.
Therefore we can focus on the pair $(a,\varphi)$. From the
simultaneous appearance of terms involving $x\partial_x$ and $x^2$
respectively we can guess that the solution to the equations of
motion will asymptotically look like a Gaussian. The corresponding
localization of the fluctuation modes around $x=0$ is consistent
with the string theory picture, where strings stretching from one
brane to the other are naturally confined to the intersection
point due to their finite tension. Setting

\begin{equation}
a_n(x)=\exp(-qx^2/2)p_n(x), \qquad
\varphi_n(x)=\exp(-qx^2/2)q_n(x) \label{Gaussian}
\end{equation}

leads to modified equations of motion:

\begin{equation}
\begin{pmatrix}
   -q^2x^2+m_n^2& -q-q^2x^2+qx\partial_x \\
  +q^2x^2-qx\partial_x-2q & m_n^2-q+q^2x^2-2qx\partial_x+\partial_x^2
\end{pmatrix}
\begin{pmatrix}
  p_n(x) \\
  q_n(x)
\end{pmatrix}=0.
\label{matrixequation2}
\end{equation}

If we assume that $p_n$ and $q_n$ are polynomials, this implies
that the fluctuation wave functions are normalizable indeed.
Polynomial solutions are possible only for discrete values of
$m^2_n$. These are given by

\begin{equation}
m^2_n=(2n-1)q;\quad n=0,2,3,4,\dots \label{massvalues}
\end{equation}

In addition, there are solutions for $m^2=0$, which were
omitted in ref. \cite{Hashimoto03}. The mass values
\eqref{massvalues} are consistent to the worldsheet analysis
\eqref{lowlyingstringspectrum} up to a substitution
$\tan(\theta/2)\rightarrow\theta/2$. This discrepancy reflects the
fact that the Yang-Mills description is only viable for small
angles. The corresponding solutions for the fluctuation wave
functions are:

\begin{align}
p_n(x)&=-\sum_{j=1,3,\cdots}^{n}(-1)^{j/2}\frac{4^{j/2}}{j!}\frac{n(n-2)\cdots(n-j+2)}{2n-1}(j-1)\left(x\sqrt{q/2}\right)^j\\
\label{pn}q_n(x)&=\sum_{j=1,3,\cdots}^{n}(-1)^{j/2}\frac{4^{j/2}}{j!}\frac{n(n-2)\cdots(n-j+2)}{2n-1}(2n-j-1)\left(x\sqrt{q/2}\right)^j
\end{align}

for $n=0,2,4,\dots$ and

\begin{align}
p_n(x)&=-\sum_{j=1,3,\cdots}^{n}(-1)^{j-1/2}\frac{4^{j-1/2}}{j!}
\left(\frac{j-1}{2}\right)(n-3)\cdots(n-j+2)\left(x\sqrt{q/2}\right)^j\\
q_n(x)&=\sum_{j=1,3,\cdots}^{n}(-1)^{j-1/2}\frac{4^{j-1/2}}{j!}
\left(n-\frac{j+1}{2}\right)(n-3)\cdots(n-j+2)\left(x\sqrt{q/2}\right)^j
\end{align}

for $n=3,5,7,\dots$. These solutions were given by Hashimoto and
Nagaoka in \cite{Hashimoto03}. A curious fact is that there is no
polynomial solution for $n=1$; $m^2=q$ although this mass level
has to be included from the worldsheet perspective. The solutions
for $(b,\psi)$ are obtained by

\begin{equation}
b_n(x)=\exp(-qx^2/2)p_n(x), \qquad \psi_n(x)=-\exp(-qx^2/2)q_n(x).
\end{equation}

Hashimoto and Nagaoka pointed out that the geometric
interpretation of the tachyonic modes is a brane recombination
process. The negative mass squared means that a non-zero tachyon
amplitude blows up exponentially

\begin{equation}
\varphi(t,x)=C(t)\varphi(x)
\end{equation}

with

\begin{equation}
C(t)=\exp(\sqrt{-m^2}t)C(0)=\exp(\sqrt{q}t)C(0)
\end{equation}

and similarly for $a$, $b$, and $\psi$. Now turn on the tachyonic
mode of one pair, say $(a,\varphi)$ and consider the total scalar
field $\Phi(x)$, including background and fluctuations, with the
explicit expression for the tachyonic mode plugged in:

\begin{align}
\Phi(t,x)&=\Phi^0(x)+\varphi(t,x)\frac{1}{2}\tau^1\\
&=\frac{1}{2}\begin{pmatrix}
  qx & C(t)\varphi(x) \\
  C(t)\varphi(x)& -qx
\end{pmatrix}
\end{align}

To get to a geometric interpretation, one has to gauge transform
$\Phi(x)$ such that it becomes diagonal. Under a gauge
transformation $U$, $\Phi$ transforms as $\Phi\rightarrow U\Phi
U^{-1}$ (remember that $\Phi$ is derived from the gauge field
$A_2$ by dimensional reduction), so that we are faced with an
ordinary eigenvalue problem which is easily solved. The
diagonalized field is given by

\begin{equation}
\label{diagphi} \Phi(t,x)=\frac{1}{2}\begin{pmatrix}
  \sqrt{q^2x^2+C(t)^2\varphi(x)^2} & 0 \\
  0 & -\sqrt{q^2x^2+C(t)^2\varphi(x)^2}
\end{pmatrix}.
\end{equation}

The corresponding brane coordinates are recovered by setting
$Y_i=2\pi\alphaprime\Phi_{ii}$. Using the explicit expression for
the tachyonic mode [cf. \eqref{Gaussian} and \eqref{pn}], they are
given by

\begin{equation}
Y_{1,2}=\pm \pi\alphaprime \sqrt{q^2x^2+C(t)^2exp(-qx^2/2)}
\label{branecoordinates}
\end{equation}

\begin{figure}
  \centering
  \includegraphics[width=12cm]{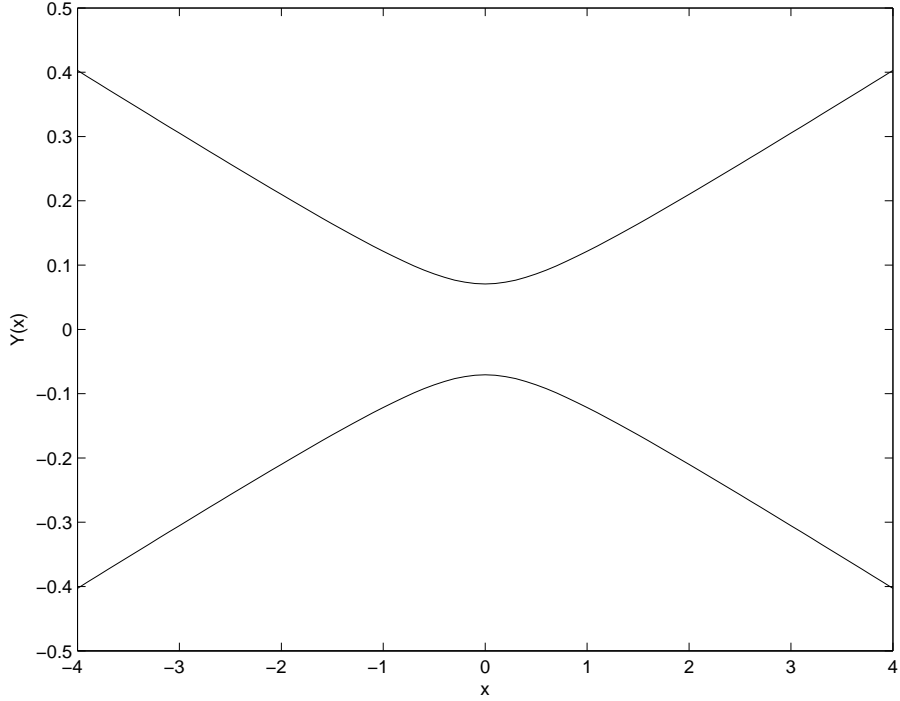}
  \caption{Recombination of intersecting D-Strings via tachyon condensation.
  For this plot, we have evaluated equation \eqref{branecoordinates} with parameters $q=1/10$ and $C^2=1/200$.
  $\pi\alphaprime$ has been set to unity.}
  \label{branerecombination}
\end{figure}

This clearly describes a recombination process, which is
illustrated in figure \ref{branerecombination}.

\subsection{Moduli in intersecting branes} \label{secmoduli}

As we mentioned before, there is an infinite family of moduli in
addition to the tachyonic modes and the tower of massive states
that were discussed in the last paragraph. These were overlooked
by Hashimoto and Nagaoka. With the same conventions as used in the
last section, setting $m^2=0$ gives the following equations of
motion for the polynomial factors $(p,q)$ of the $(a,\phi)$ pair
of fluctuation fields:

\begin{equation}
\begin{pmatrix}
   -q^2x^2 & -q-q^2x^2+qx\partial_x \\
  +q^2x^2-qx\partial_x-2q & -q+q^2x^2-2qx\partial_x+\partial_x^2
\end{pmatrix}
\begin{pmatrix}
  p_n(x) \\
  q_n(x)
\end{pmatrix}=0.
\end{equation}

The dependence on $q$ can be eliminated by rescaling $x\rightarrow
\sqrt{q/2}\ x$. Then the equations take the following form

\begin{equation}
\begin{pmatrix}
   -2x^2 & -1-2x^2+x\partial_x \\
  2x^2-x\partial_x-2 & -1+2x^2-2x\partial_x+\frac{1}{2}\partial_x^2
\end{pmatrix}
\begin{pmatrix}
  p_n(x) \\
  q_n(x)
\end{pmatrix}=0.
\end{equation}

To solve these equations, we define new polynomials $f$, $g$ by

\begin{align}
f(x)&=\frac{1}{2}(p(x)+q(x))\\
g(x)&=\frac{1}{2}(p(x)-q(x))
\end{align}

leading to equations of motion:

\begin{equation}
\begin{pmatrix}
  -4x^2-1+x\partial_x & 1-x\partial_x \\
  -8x^2-\partial_x^2+6x\partial_x+6 & -2x\partial_x+2+\partial_x^2
\end{pmatrix}
\begin{pmatrix}
  f(x) \\
  g(x)
\end{pmatrix}=0.
\end{equation}

If one defines

\begin{equation}
f(x)=\sum_{j=0}^{n}f_j x^j;\qquad g(x)=\sum_{j=0}^{n'}g_j x^j
\end{equation}

with $g_n\equiv 1$, one gets equations for the coefficients:

\begin{align}
j\geq 2:&
 \begin{cases}
 -(j-1)g_j+(j-1)f_j-4f_{j-2}=0 \\
 (j+2)(j+1)g_{j+2}-2(j-1)g_j\\
 \quad -(j+2)(j+1)f_{j+2}+6(j+1)f_j-8f_{j-2}=0
 \end{cases}\\
j=1:& \quad g_3-f_3+2f_1=0\\
j=0:&
 \begin{cases}
 g_0-f_0=0\\
 g_2-f_2+4f_0=0
 \end{cases}
\end{align}

While setting $j=n+2$ shows that $f_n=0$, setting $j=n$ leads to
$f_{n-2}=-(n-1)/4$. Now setting $j=n-2$ gives two equations for
the two variables $f_{n-2}$ and $g_{n-4}$. It turns out that these
are not linearly independent, so one can freely choose one of the
variables. It is convenient to require $f_{n-4}=0$, which means
that we must have $f_{n-2}=g_{n-2}$. Then all other component
equations can be satisfied by setting $f_j=g_j=0$ for $j\leq n-2$.
Because we chose to set $f_{n-4}=0$, the above solution is not
unique but note that giving any other value to $f_{n-4}$ is
equivalent to adding a solution of the form $\{g_{n-2}=1$, $g_j=0$
for $j>(n-1)\}$. Altogether we have solutions

\begin{equation}
\begin{pmatrix}
  f(x) \\
  g(x)
\end{pmatrix}=
\begin{pmatrix}
  -\frac{n-1}{4}x^{n-2}\\
  x^n-\frac{n-1}{4}x^{n-2}
\end{pmatrix}
\end{equation}

for $n=2,3,\dots$ and, in addition, the solution $(f,g)=(0,x)$.
The most general solution is given by a linear combination of
these. In terms of the original variables, the solutions are given
by

\begin{equation}
\label{masslesssolutions1}
\begin{array}{ccl}
  a(x) & = &\exp\left(-qx^2/2\right)(qx^n-(n-1)x^{n-2}),\\
  \varphi(x) & = &-\exp\left(-qx^2/2\right)qx^n
\end{array}
\end{equation}

for $n=2,3,\dots$ where we absorbed a global ($n$-dependent)
factor by rescaling. The additional solution takes the form

\begin{equation}
\label{masslesssolutions2}
\begin{array}{ccl}
a(x)&=&\exp\left(-qx^2/2\right)x,\\
\varphi(x)&=&-\exp\left(-qx^2/2\right)x.
\end{array}
\end{equation}

As before, the solutions for $(b,\psi)$ can be obtained by
substituting $(a,\phi)\leftrightarrow(b,-\psi)$. To see the
geometrical interpretation of the zero modes, we refer to the
diagonal form \eqref{diagphi} of the scalar field $\Phi$, with
fluctuations included. The worldlines of the D-strings are given
by

\begin{equation}
Y(t,x)=\pm 2\pi\alphaprime
\frac{1}{2}\sqrt{q^2x^2+C(t)\varphi(x)^2}.
\end{equation}

The zero mass modes do not evolve in time so that $C$ is simply a
constant. For $\varphi(x)$ we can take any superposition of the
massless solutions \eqref{masslesssolutions1} and
\eqref{masslesssolutions2}, so that the most general case is given
by $\varphi(x)=x\exp(-qx^2/2)P(x)$ with $P(x)$ an arbitrary
polynomial. This means that we can arbitrarily deform the
worldlines of the D-strings in the vicinity of the intersection
point as long as the deformations remain small (away from the
intersection point, the fluctuations are damped by the exponential
factor). There is one restriction to possible deformations, namely
that because $\varphi(x)$ goes to zero like $x$ or faster when $x$
approaches zero, the worldlines will always intersect.
Intuitively, deformations of the D-string worldlines should not be
massless because the D-strings are wrapped around a torus
(possibly with infinite radius) and they have a finite tension
which should work against any wriggles in the worldlines. But note
that the tension is not the only contribution to the energy
density because $\varphi$ can only be turned on in conjunction
with $a$, which also contributes to the $D_{\mu}\Phi D^{\mu}\Phi$
term. Since massless fluctuation modes by construction do not
change the energy of a background configuration, we have to assume
that all contributions to the energy coming from fluctuation modes
cancel in the end. The question remains what the meaning of the
massless modes is, in particular because they do not appear in the
string theory spectrum. The logical answer to this is that the
massless modes are remnants of the original gauge symmetry.
Turning on their amplitude is equivalent to gauge transforming the
background fields. To see this, we consider a gauge transformation
of the form

\begin{align}
A_1^0 & \rightarrow \tilde{A}^0_1\equiv UA_1^0U^{-1}-(\partial_xU)U^{-1}\nonumber\\
\Phi^0 & \rightarrow \tilde{\Phi^0}\equiv
U\Phi^0U^{-1}\label{transformedbackground}
\end{align}

where

\begin{align}
U(x)&=\exp(i\Lambda(x))\nonumber\\
\Lambda(x)&=\exp(-qx^2/2)x^{n-1}(\tau_2/2).
\end{align}

If we consider infinitesimal gauge transformations, we can neglect
quadratic and higher terms in $\Lambda$. By plugging in the
explicit expressions \eqref{YMbackground} for the intersecting
brane background into equations \eqref{transformedbackground} and
expanding to linear order in $\Lambda$ we find that the
transformed background fields are given by

\begin{align}
\tilde{A}^0_1&=A^0_1+\exp(-qx^2/2)(qx^n-(n-1)x^{n-2})(\tau_2/2)\\
\tilde{\Phi}^0&=\Phi^0-\exp(-qx^2/2)qx^n(\tau_1/2).
\end{align}

As we expected, these expressions are the sum of the untransformed
background fields and fluctuation fields of the form
\eqref{masslesssolutions1}. In other words, the massless
fluctuation modes generate gauge transformations of the background
fields.\footnote{Gauge transformations with $\Lambda\propto\tau_1$
are generated by massless fluctuations of the $(b,\psi)$-fields.}
In fact, choosing a particular gauge background breaks the
$U(2)$-gauge symmetry of the Yang-Mills action spontaneously and
the massless fluctuations are components of the corresponding
Goldstone scalar. Since gauge transformations relate physically
equivalent configurations, this Goldstone scalar describes a
redundancy of the formalism rather than a physical degree of
freedom. Thus, we resolve the apparent contradiction to the string
worldsheet analysis.


\section{Branes at large angles}

As has been pointed out, non-Abelian Yang-Mills theory fails to
describe intersecting branes at large angles. However, by
increasing one intersection angle continuously (keeping all other
angles at zero) one eventually arrives at a coincident
brane-antibrane pair. This is a configuration which has been
thoroughly investigated from the tachyon field theory perspective
and it seems promising to generalize existing tachyon actions to
include small intersection angles between the brane and antibrane.

Much of the work on the brane-antibrane system has been inspired
by Sen's conjectures on tachyon condensation \cite{Sen98a, Sen98b,
Sen99a, Sen99b}. With the tachyon rolling down towards its
potential's minimum, standard first quantized string theory fails
because it is only defined around the perturbative vacuum.
Therefore one has to resort to string field theory methods. For
some time the focus had been on numerical studies using Witten's
cubic string field theory but more recently Boundary String Field
Theory (BSFT) has produced a number of exact results on tachyon
condensation. One particular success of BSFT has been the
derivation of an effective tachyon action for a non-BPS brane.

\begin{equation}
\label{nonBPSaction} S=-\int d^{p+1}x
e^{-T^2/4}\left(1+\frac{1}{2}\partial_{\mu}T\partial^{\mu}T\right)
\end{equation}

This action had originally been proposed as a toy model for
tachyon condensation \cite{Minahan00a} and was subsequently shown
to be an exact two-derivative truncation of the full BSFT
\cite{Kutasov00b}.\footnote{In fact, this is not quite correct.
The action which was derived from BSFT in \cite{Kutasov00b} has a
different numerical coefficient for the kinetic term, leading to
the wrong tachyon mass at $T=0$. This is a well known but as of
yet unresolved discrepancy. In \eqref{nonBPSaction}, it has been
corrected by hand.} We are not keeping track of overall numerical
factors of the action here and $\alphaprime$ has been set to
unity.

A non-BPS brane is a ``wrong p'' Dp-brane meaning odd dimension
D-branes in type IIA theory and even dimension D-branes in type
IIB theory. It breaks supersymmetry completely and is expected to
decay to a stable BPS D(p-1)-brane via tachyon condensation. We
will use the non-BPS brane tachyon action as a starting point for
establishing an effective action for branes and antibranes
intersecting at small angles. There will be two major
modifications to the action \eqref{nonBPSaction}. First, it has to
be lifted to the brane-antibrane case, which is in fact closely
related to the non-BPS brane via brane descent relations. Next, we
need to introduce gauge fields to the action because a non-zero
magnetic flux is related to non-zero intersection angles via
T-duality. In the BSFT approach, it is notoriously difficult to
include gauge fields from the beginning because the corresponding
boundary terms introduce non-trivial interactions in the BSFT
action. Thus, the theory is no longer exactly solvable and one has
to make do with perturbational computations. However, one can
sidestep these complications by manufacturing a tachyon action
``by hand''. We will explain the sort of hand-waving arguments
which are needed in such a construction and ``derive'' a simple
gauged $D\overline{D}$-action.

\subsection{A simple gauged $D\overline{D}$-action}

In the brane-antibrane system, the tachyonic modes come from
strings which connect brane and anti-brane. The fact that there
are two possible orientations for such strings leads to a
degeneracy of the string theory spectrum. This can be dealt with
by making the tachyon field in \eqref{nonBPSaction} complex. Thus,
a natural generalization of the non-BPS-brane action would be:

\begin{equation}
\label{DDbarlowenergy} S=-\int d^{p+1}x
e^{-\overline{T}T/4}\left(1+\frac{1}{2}\partial_{\mu}\overline{T}\partial^{\mu}T\right).
\end{equation}

This action has global $U(1)$-symmetry. To gauge the symmetry in a
standard way, the derivatives have to be covariantized,
$\partial_{\mu}\rightarrow D_{\mu}$ where

\begin{equation}
\label{tachyoncovariant} D_{\mu}=\partial_{\mu}-iA_{\mu}.
\end{equation}

What is the meaning of the gauge field $A_{\mu}$, which we just
introduced? Clearly, it should be related to some abelian subgroup
of the original $U(2)$-symmetry of a two-brane system. The
Chan-Paton representation of the complex tachyon state is given by

\begin{equation}
\begin{pmatrix}
  0 & \overline{T} \\
  T & 0
\end{pmatrix}.
\end{equation}

This transforms under the adjoint representation of $U(2)$. Now
consider a special family of $U(2)$ transformations,

\begin{equation}
\begin{pmatrix}
  0 & \overline{T} \\
  T & 0
\end{pmatrix}\rightarrow
\begin{pmatrix}
  0 & \rho_1^*\rho_2\overline{T} \\
  \rho_2^*\rho_1T & 0
\end{pmatrix}=
\begin{pmatrix}
  \rho_1^* & 0 \\
  0 & \rho_2^*
\end{pmatrix}
\begin{pmatrix}
  0 & \overline{T} \\
  T & 0
\end{pmatrix}
\begin{pmatrix}
  \rho_1 & 0 \\
  0 & \rho_2
\end{pmatrix}
\end{equation}

where $\rho_1^*\rho_1=\rho_2^*\rho_2=1$. Clearly, the tachyon
state is neutral under the combination $\rho_1=\rho_2$ while it is
charged under $\rho_1=\rho_2^*$. Since $\rho_1$ and $\rho_2$
correspond to abelian gauge transformations on the first and
second brane respectively, the correct form for the covariant
derivative in \eqref{tachyoncovariant} is given by
$A_{\mu}=A_{\mu}^{(1)}-A_{\mu}^{(2)}$ where $A_{\mu}^{(i)}$ are
the abelian gauge fields on the individual branes. Next, we have
to include gauge invariant kinetic terms for the gauge field in
the tachyon action. The simplest option clearly is:

\begin{equation}
S=-\int d^{p+1}x
e^{-\overline{T}T/4}\left(1+\frac{1}{4}F_{\mu\nu}F^{\mu\nu}+\frac{1}{2}D_{\mu}\overline{T}D^{\mu}T\right).
\end{equation}

Introducing kinetic terms in this way also ensures that for
$T\equiv0$ the theory reduces to standard gauge theory of the
unbroken Abelian symmetry. Of course, this is by far not the only
possibility for including gauge kinetic terms. We will shortly
discuss two more options at the end of this section. For now, we
will stick to the simple action given above. Being interested in
low-energy physics near the perturbative vacuum at $T=0$, we can
make a further approximation. After reinserting factors of
$\alphaprime$, the action becomes

\begin{equation}
\label{simpleaction} S=S_0-\int d^{p+1}\left(\frac{1}{4}
F^{\mu\nu}F_{\mu\nu}+\frac{1}{2}
D_{\mu}\overline{T}D^{\mu}T-\frac{1}{4\alphaprime}
\abs{T}^2\right)
\end{equation}

up to zeroth order in $\alphaprime$ ($\cong$ low-energy) and
second order in the tachyon field ($\cong$ near the perturbative
vacuum). Note that since we are not keeping track of overall
numerical factors, the order in $\alphaprime$ is only relative.
>From this, the action for intersecting brane-antibrane systems
(i.e. intersecting branes at large angle) is derived by applying
T-duality in the form of dimensional reduction. The part of the
action containing the tachyon field becomes:

\begin{gather}
\label{intersectingDDbar} S=-\int d^{p'+1}
\left(\partial_{\mu}\overline{T}\partial^{\mu}T+
\left(\sum_{I}\frac{\Delta Y^I(x)\Delta
Y^I(x)}{(2\pi\alphaprime)^2}-\frac{1}{2\alphaprime}\right)\abs{T}^2\right)\\
\Delta Y^I\equiv Y^{(1)I}-Y^{(2)I}
\end{gather}

where $Y^{(i)I}$ are the transverse coordinates of brane ($i=1$)
and antibrane ($i=2$) and we have set the gauge field on the
remaining brane dimensions to zero. Now we see that everything
fits together neatly. For parallel brane and antibrane with
non-zero separation, the term with the $\Delta Y$'s simply
reproduces the tension from stretched strings which reduces the
negative mass squared of the tachyon. For intersecting branes, the
same term leads to a localization of the tachyon modes on the
intersection manifold and a mass spectrum with towers of evenly
spaced states. For simplicity, we consider the string-antistring
case. Then, the background corresponding to an intersection angle
$\phi$ is given by $Y^{(1,2)}=\pm\tan(\phi/2)x$ and therefore
$\Delta Y(x)=2\tan(\phi/2)x$. The action \eqref{intersectingDDbar}
is by virtue of its ``derivation'' essentially a small angle
approximation because $\tan(\phi/2)$ is proportional to the field
strength amplitude in \eqref{simpleaction} and higher order terms
in the field strength would correspond to corrections which are
higher order in $\alphaprime$. Thus, within the range of validity
of \eqref{intersectingDDbar} we can approximate by setting
$\tan(\phi/2)\approx\phi/2$. The action becomes

\begin{equation}
\label{DstringDantistring} S=\int
dtdx\left[\partial_t\overline{T}\partial_tT-\overline{T}\left(-\partial_x^2
+\left(\frac{\phi}{2\pi\alphaprime}\right)^2x^2-\frac{1}{2\alphaprime}\right)T\right].
\end{equation}

This involves a harmonic oscillator potential. Thus, the tachyon
fluctuation modes are localized at the intersection point at
$x=0$. The mass spectrum of the tachyon field fluctuations which
one computes from the above action is given by

\begin{equation}
m^2=-\frac{1}{2\alphaprime}+\frac{\phi}{2\pi\alphaprime}(2n+1),\quad
n=0,1,2,\dots
\end{equation}

Surprisingly, this coincides \emph{exactly} with the worldsheet
analysis which was given in section \ref{SecWS} [cf.
\eqref{DDbarspectrum}]. This observation stays valid for higher
dimensional branes and multiple intersection angles. The exact
matching of mass spectra should be seen as a strong argument in
favor of the empirical action \eqref{simpleaction}.

\begin{figure}
  \centering
  \includegraphics[width=12cm]{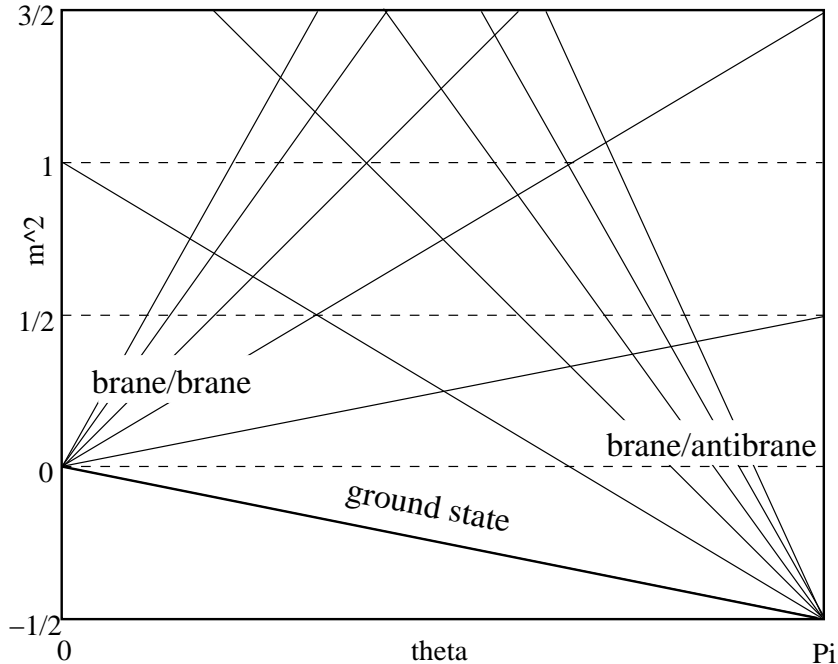}
  \caption{The ground state of the tower of lowest mass states at brane-brane configurations flows
  to the ground state of the tower of lowest mass states at brane-antibrane configurations, where the
  intersection angle $\theta$ flows from $0$ to $\pi$. The excited states of the individual towers do
  not match.}\label{figmatchstates}
\end{figure}

In \cite{Hashimoto03}, Hashimoto and Nagaoka used the action
\eqref{intersectingDDbar} in their discussion of large brane-brane
intersection angles. However, they compared the fluctuation
spectrum to the mass spectrum formula
$m^2=-\theta/2\pi+n\theta/\pi,\quad n=0,1,2\dots$ This formula is
routinely used in many publications on intersecting branes and can
be derived from a worldsheet analysis. However, as we pointed out
in section \ref{SecWS}, it only represents the lowest energy
states for small angles. Thus, they could not match the
fluctuation mass spectrum to the worldsheet analysis.
Coincidentally, the lowest mass value is the same in both cases
and therefore their main point, which was only concerned with the
lowest mass state, stays valid. For an illustration of this point,
see figure \ref{figmatchstates}.

\subsection{More tachyon actions}

We will now shortly discuss other proposals for effective tachyon
actions on the brane-antibrane system. An attractive option, which
is reminiscent of the Abelian Born-Infeld action, is the
following:

\begin{equation}
\label{tachyonBI} S=-\int d^{p+1}x
V(\abs{T})\sqrt{-\det(\eta_{\mu\nu}+2\pi\alphaprime F_{\mu\nu}+
D_{\{\mu}\overline{T}D_{\nu\}T})}
\end{equation}

with unspecified tachyon potential $V$. This is actually a
generalization of a non-BPS-brane action which had been proposed
in \cite{Garousi00} and was shown to reproduce (in the
non-BPS-brane case) some S-matrix elements involving tachyon
states. The Born-Infeld like tachyon action \eqref{tachyonBI} and
similar actions have been used as phenomenological field theory
models of tachyon condensation \cite{Gibbons00, Sen02c}. Although
it is quite different from the simple action we have discussed in
the previous section, its low-energy limit near the perturbative
vacuum actually is the same. Expanding the determinant, using the
standard formula $\frac{\partial}{\partial A_{ij}}\det A=\det A
A^{-1}_{ij}$, we arrive at

\begin{equation}
S=-\int d^{p+1}x
V(\abs{T})\left(1+\frac{1}{2}G^{\mu\nu}D_{\mu}\overline{T}D_{\nu}T
+\mathcal{O}(\abs{DT}^4) \right)
\end{equation}

where $G=(\eta+2\pi\alphaprime F)^{-1}_{\text{sym}}$ is the open
string metric. Expanding the potential as
$V(\abs{T})=V_0(1+a\abs{T}^2)+\mathcal{O}(\abs{T}^4)$ and
discarding all quartic terms in the tachyon field and its
derivatives, the action becomes

\begin{equation}
S=-\int d^{p+1}x
\left(a\abs{T}^2+\frac{1}{2}G^{\mu\nu}D_{\mu}\overline{T}D_{\nu}T\right)
\end{equation}

up to constant terms and overall numerical factors. Setting the
gauge fields to zero, the coefficient $a$ is determined by the
requirement that the tachyon field should have
$m_T^2=-1/2\alphaprime$. Thus, we have $a=-1/4$. In the low-energy
limit, the open string metric $G^{\mu\nu}$ reduces to the standard
Lorentzian metric and by T-dualizing we recover action
\eqref{intersectingDDbar} of the previous section. The discussion
of the mass spectrum is identical. In this light, it seems that
the mass spectrum matching for intersecting brane-antibrane is a
universal feature of brane-antibrane actions rather than being
model-dependent. It simply is a consequence of the specific gauge
symmetry of the system and the input of the correct tachyonic mass
at zero angle.

There are obviously many more ways of producing gauged
brane-antibrane actions. Notably, Sen recently proposed a
brane-antibrane action which is related to the Born-Infeld-type
action \eqref{tachyonBI} but takes a particularly intuitive form
\cite{Sen03}:

\begin{equation}
\label{Sensaction} S=-\int d^{p+1}x V(T,Y_{(1)}^I-Y_{(2)}^I
)\left(\sqrt{-\det\mathbf{A}_{(1)}}+\sqrt{-\det
\mathbf{A}_{(2)}}\right)
\end{equation}

where

\begin{gather}
\mathbf{A}_{(i)\mu\nu}=\eta_{\mu\nu}+2\pi\alphaprime F_{\mu\nu}^i+\partial_{\mu}Y^I_{(i)}\partial_{\nu}Y^I_{(i)}+D_{\{\mu}T^*D_{\nu\}}T\\
F_{\mu\nu}^{(i)}=\partial_{\mu}A^{(i)}_{\nu}-\partial_{\nu}A^{(i)}_{\mu},\quad
D_{\mu}T=(\partial_{\mu}-i(A_{\mu}^{(1)}-A_{\mu}^{(2)}))T
\end{gather}

and $A^{(i)}_{\mu}$ are the Abelian gauge fields on the brane
($i=1$) and the antibrane ($i=2$) respectively. Other than in
previous examples, the transverse brane coordinates $Y_{(i)}^I$
are included from the beginning. The motivation for their
inclusion in \cite{Sen03} was to allow for non-zero separation of
a parallel brane-antibrane pair rather than to include the
intersecting brane-antibrane case. In expanding the tachyon
potential $V$ to quadratic order in $\abs{T}$, the coefficients of
the expansion are determined by the requirements that

\begin{itemize}
\item[1.] For $T=0$, the action should reduce to the sum of the
actions on the two individual branes.
\item[2.] The tachyon mass should conform with the value from the
string worldsheet computation for a parallel brane-antibrane pair
with non-zero separation
\end{itemize}

These requirements lead to

\begin{equation}
\label{Senspotential}
V(T,Y^I_{(1)}-Y^I_{(2)})=\tau_p\left[1+\frac{1}{2}
\left\{\sum_I\left(\frac{Y^I_{(1)}-Y^I_{(2)}}{2\pi\alphaprime}\right)^2
-\frac{1}{2\alphaprime}\right\}\abs{T}^2+\mathcal{O}(\abs{T}^4)\right]
\end{equation}

where $\tau_p$ is the tension of the individual Dp-branes. Thus,
by expanding the square roots in \eqref{Sensaction} to lowest
order in the tachyon fields and taking the low-energy limit, we
once more recover the action \eqref{intersectingDDbar} of the
previous section. This is a further example for the universality
of mass spectrum matching on the intersecting brane-antibrane
system.


\section{Conclusion and Discussion}

In an attempt to establish a field theory description of the
brane-antibrane pair intersecting at small angles we made the case
for the following effective action:

\begin{equation}
\label{simpletachyonaction} S=-\int d^{p+1}\left(\frac{1}{4}
F^{\mu\nu}F_{\mu\nu}+\frac{1}{2}
D_{\mu}\overline{T}D^{\mu}T-\frac{1}{4\alphaprime}
\abs{T}^2\right)
\end{equation}

This action is understood as a low-energy approximation of the
brane dynamics near the perturbative vacuum. Its mass spectrum was
shown to coincide perfectly with the string theory worldsheet
analysis. Different generalizations of tachyon field theories to
intersecting brane-antibranes were discussed and we indicated that
due to the specific gauge symmetry of the system, their low-energy
limit had to be the same in all cases. It would be interesting to
undertake a perturbational BSFT computation with gauge fields
included from the beginning. For attempts at this consider
\cite{Kraus00, Takayanagi00}. In particular, the authors of
\cite{Takayanagi00} claimed to have performed computations up to
order $\alphaprime^2$. It would be interesting to check their
results (which effectively provide higher order terms in the
intersection angles) against the worldsheet analysis of the string
theory mass spectrum.

Local brane recombination via tachyon condensation has been
proposed as a realization of the Higgs Effect in brane world
models. Our analysis suggests that
starting from a stable, tachyon-free D-brane configuration
and then rotating
some of the D-branes
by a small angle, tachyon condensation can indeed be
equivalently described by the standard Higgs effect in an effective
gauge theory.
However the case of rotating D-branes
in an unstable brane-antibrane pair by a small
angle is more problematic for  phenomenological purposes, since
in the brane-antibrane system
all open string degrees of freedom disappear after the tachyon condensation.
This includes in particular also
the open string gauge field excitations, which means
in field theory language that the gauge bosons become
infinitely heavy at the end point of the tachyon condensation.
Therefore this kind of open string gauge bosons in the
brane-antibrane system cannot play the role of the weak vector
bosons in the standard model.
%
%
%
In \cite{Hashimoto03}, Hashimoto and Nagaoka showed
how the simple tachyon action \eqref{simpletachyonaction} could be
used to realize local brane-antibrane annihilation through a
backreaction from the tachyon field to the brane coordinates.
However, in their analysis they assumed that only the ground state
would condense. The role of higher level tachyonic modes in brane
recombination/local brane-antibrane annihilation remains
unexplored.

Finally, it might also be interesting to explore the consequence
of the large number of tachyonic modes for brane and antibrane
intersecting at small angles to inflation models in brane
cosmology.

After finishing the paper we noticed the paper \cite{Jones03},
where an expression for the tachyon potential for intersecting
branes at arbitrary angles was derived in the context of boundary
superstring field theory. We are grateful to N. Jones for drawing
our attention to this work.

\bibliographystyle{elsevier}


\end{document}